\documentclass[12pt,preprint]{aastex}

\shorttitle{Compositions of Stars with Planets}
\shortauthors{Gonzalez}

\begin{document}

\title{The Chemical Compositions of Stars with Planets: \\
    A Review}

\author{Guillermo Gonzalez}
\affil{Department of Physics and Astronomy, Iowa State University,
    Ames, IA 50011}
\email{gonzog@iastate.edu}

\begin{abstract}
A number of trends among the properties of exoplanets have become evident in the years since the first one was announced in 1995. One particularly interesting trend began to emerge in 1997 -- the incidence of giant planets correlates with the metallicity of the host star. This has since been established with a high degree of statistical significance by several research groups. Other, more subtle, trends are beginning to appear as the sample size continues to grow and the statistics improve. I review the state of our knowledge concerning the observed compositional trends and their possible causes and suggest several research directions.
\end{abstract}


\section{Introduction}

The announcement of the first planetary mass object orbiting another sun-like star by \citet{mayq95} opened a new field of empirical study to astronomers. At the time of this writing (August 2006), nearly 200 exoplanets have been reported. Nearly all have been found with the Doppler method, whereby the reflex motion of the host star about the system's center-of-mass is measured over at least one orbital period. A few have also been found with the transit and microlensing methods.

Several global trends have been found in the database of extrasolar planets around nearby FGK stars \citep{marcy05, but06}: 1.2 per cent of the planets are found in orbits with semi-major axis, $\it a$, less than 0.1 AU; the distribution of planet mass is a strongly decreasing function of planet minimum mass, $\it M$,  ($dN/dM \propto M^{-1.1}$) with a sharp cutoff above 12 M$_{\rm J}$; $> 7$ percent of surveyed stars have planets with semimajor axis, $\it a$, $< 5$ AU; eccentric orbits are common, with a median eccentricity value of 0.25; there is a paucity of high-mass planets with short-period orbits. 

These discoveries have spurred much theoretical work. Some have focused on explaining the high eccentricities (see review by \citet{trem04}), while others have explored mechanisms of planet migration to smaller orbits (see review by \citet{thom05}). These studies are based on one of two broad categories of planet formation theories: core instability accretion (CIA, \citet{poll96}); disk gravitational instability (DGI, \citet{boss04} and references therein). The CIA model is more mature, and it has had some success in explaining some of the observed trends (e.g., \citet{idalin05}).

The topic of the present review, however, concerns the compositions of the planet host stars. \citet{gg97} first showed a link between the metallicities of the host stars and the presence of planets. Many studies have since been published on this topic, the most recent ones confirming the link with a high degree of statistical significance (e.g., \citet{sant05,fv05}). In this review I will describe research on the composition of stars with planets (SWPs), focusing on studies published since my \citet{gon03} review. In \S~2 I summarize the ongoing exoplanet surveys and the spectroscopic observations and analyses of SWPs. I briefly review in \S~3 survey statistics, including the preparation of control samples, biases and the incidence of exoplanets. In \S~4 I summarize the results of recent spectroscopic chemical abundance analyses of SWPs, and in \S~5 I discuss the proposed explanations for the observed abundance anomalies among SWPs. I end in \S~6 by suggesting several research projects that may help advance our understanding of planet formation and evolution by studying the chemical compositions of SWPs.

\section{Exoplanet Surveys}

Currently, three distinct types of surveys are yielding exoplanet candidates. The Doppler method is the oldest and most successful one. Most stars in the Doppler surveys are selected from the brightest F to M spectral type dwarf and subgiant stars in the solar neighborhood, typically limited to about 70 parsecs distance; the samples are magnitude limited, not volume limited. To date, most SWPs have been found by two groups: the California \& Carnegie and Anglo-Australian Planet Searches led by Geoff Marcy and Paul Butler\footnote{see http://exoplanets.org.}, and the Geneva Extrasolar Planet Search Programmes led by Michelle Mayor.\footnote{see http://obswww.unige.ch/~udry/planet/planet.html.} The American group is surveying 1330 stars \citep{marcy05,but06}, while the Europeans are surveying nearly 2000, with considerable overlap between the two samples. Specialized surveys restrict their targets to M dwarf \citep{endl06}, giant \citep{hat06}, Hertzsprung gap \citep{john06}, metal-poor dwarf \citep{sozz06} or metal-rich dwarf \citep{deb05,mou06} stars.

Several planets have been either confirmed or discovered with the photometric transit method (reviewed by \citet{char06}). Transit surveys target either individual star clusters or large regions of the Milky Way. When combined with Doppler data, transit data allow determination of the true mass of a planet. Microlensing surveys have also yielded a few exoplanet candidates \citep{be06}.

Most of the giant planets have been found around F, G and K dwarf stars. Several have been found around subgiants, and a few have been found around giants. Only two M dwarfs are known to harbor planets, even though many M dwarfs are included in the surveys. 

\subsection{Spectroscopic Observations}

Most of the abundance analyses of SWPs are based on high S/N ratio, high-dispersion echelle spectra. They typically cover most or all of the optical spectrum with S/N ratios of at least several hundred 
per resolution element and resolving power near 60,000. Many SWPs have been observed by more than one group, allowing useful cross-checks of their data analysis methods. Most spectra have been obtained with 2 - 3 meter telescopes, though some have been obtained with slightly smaller or much larger telescopes (e.g., Keck 10 meter).

Most of the nearly 200 known SWPs have received detailed spectroscopic analysis. Recent large spectroscopic surveys of SWPs (and comparison stars) include \citet{bond06}, \citet{fv05}, \citet{gil06}, \citet{luck06} and \citet{takhon05}. I will discuss these studies in later sections.

\subsection{Spectroscopic Analysis Methods}

The most popular method employed to analyze the spectra of SWPs was developed over the past few decades in studies of nearby sun-like stars. It makes use of measurements of the equivalent widths (EWs) of 40 to 60 Fe I and Fe II absorption lines, together with model stellar atmospheres, to derive the four basic stellar parameters: effective temperature (T$_{\rm eff}$), surface gravity ($g$), microturbulence velocity parameter and [Fe/H]. Sometimes T$_{\rm eff}$ and $g$ are determined from photometric and parallax measurements and theoretical stellar isochrones. The analyses employ stellar atmosphere models calculated assuming local thermodynamic equilibrium (LTE) and one-dimensional (1D) geometry. The typical quoted formal uncertainty in the derived value of [Fe/H] is 0.05 dex, but the formal uncertainty can be as low as 0.02 dex for an old G dwarf star. Strictly differential spectroscopic analyses of pairs of stars of similar spectral types can achieve formal uncertainties below 0.01 dex (e.g., \citet{lawgg01,tak16}). Typical mean differences in T$_{\rm eff}$ and [Fe/H] among spectroscopic studies of SWPs are 40-70 K and 0.01-0.02 dex, respectively (see \citet{luck06} for detailed comparisons).

There is presently some controversy regarding the absolute stellar temperature scale. This is important to abundance determinations, because the derived abundances are particularly sensitive to the assumed value of T$_{\rm eff}$. Spectroscopic studies of SWPs determine T$_{\rm eff}$ from Fe I line excitation equilibrium. The Infrared Flux Method (IRFM) uses the infrared flux relative to the total flux to determine T$_{\rm eff}$. \citet{ram05} applied the IRFM method to a sample of SWPs, deriving T$_{\rm eff}$ values about 100 K lower than studies using Fe I excitation equilibrium. This difference is highly significant. \citep{casa06}, however, derived a different calibration of T$_{\rm eff}$ using the IRFM method, obtaining very good agreement with published spectroscopic T$_{\rm eff}$ determinations. These authors trace the discrepancies in the prior studies to use of Vega as a flux standard. While the latest data favor the spectroscopic temperature scale, more research is required to settle this controversy. For example, angular diameter measurements of G dwarfs would be helpful.

The most critical atomic input parameter for a given spectral line is its oscillator strength. Two approaches are available with regard to its source. Either it is obtained from laboratory experiments or from the solar spectrum (for which a specific set of elemental abundances is adopted). To date, spectroscopic studies of SWPs have employed solar oscillator strengths; in effect, chemical abundance determinations for SWPs are differential with respect to the Sun. This is the better choice, because SWPs are similar to the Sun (in fact, the Sun is near the average of the SWP spectral types), hence eliminating possible systematic errors in the oscillator strengths and minimizing deficiencies in the models and analysis methods. Such differential systematic errors are smallest for G2 dwarfs and become progressively larger for stars increasingly different from the Sun in T$_{\rm eff}$, $g$ and [Fe/H].

\citet{vf05} have developed an automated method for stellar parameter determination from high dispersion spectra. They compare synthetic spectra calculated from model atmospheres and atomic and molecular line data to observed spectra over narrow wavelength intervals and arrive at a solution iteratively. This has the advantage over the traditional method in that it allows for large numbers of stars to be analyzed with relatively little human intervention. Using this method, they determined the properties of 1040 FGK dwarfs.

Due to their very crowded spectra, solar metallicity M dwarfs have until recently lacked accurate spectroscopic chemical abundance analyses. \citet{ww05} have demonstrated that careful selection of absorption lines in spectral regions lacking molecular bands can yield reliable metallicities for cool metal-rich dwarfs. \citet{bean06}, on the other hand, have had success analyzing regions that include TiO bands.

Once a basic set of stellar atmospheric parameters are determined for a given star, the abundances of other elements can be derived from individual spectral lines. Typically, at least two or three absorption lines are measured for each of 15 to 20 elements. The abundances are derived either from the EWs or by matching synthetic and observed spectra. The quality of an abundance determination depends on the element. Light elements, such as Li, C, N, O, Na, Al, Mg and S are represented by relatively few spectral lines in sun-like stars. Reliable abundances for them require high quality spectra. In the case of an element with many spectral lines, such as Ti or Fe, one can be more selective and include only unblended lines in the analysis.

C and O are especially important elements to include in a spectroscopic analysis, as they are important opacity sources in stars and they are important in the chemistry of protoplanetary disks. In addition, they have very low condensation temperatures and provide important leverage in one of the tests for compositional anomalies described below. The forbidden lines are considered to be the best abundance indicators for these elements, but they are weak and blended, requiring high quality spectra.

Some elements have transitions that exhibit large hyperfine splitting; examples include Mn I, Cu I and Sc I. Spectral lines of these elements must be synthesized with all the hyperfine components for the most accurate results. Similarly, determining an accurate value of the $^{6}$Li/$^{7}$Li ratio (its importance is described below) requires inclusion of both the isotopic and hyperfine components in the syntheses. 

Published studies of SWPs have yet to incorporate all the advances in stellar spectroscopic analysis made over the past few years. For instance, the next generation of analyses could relax the assumption of LTE for all the elements and replace the 1D model atmospheres with 3D versions \citep{asplund05}. Solar abundances have recently been determined incorporating these improvements \citep{asp05}. 

\section{Survey Statistics}
\subsection{Control Samples}

In order to interpret the results of a survey correctly, it is necessary to place the detected SWPs within the proper context. This requires generating a comparison or ``control'' sample, the stars of which are analyzed in the same way as the SWPs. When SWPs are compared to such a sample, several possible systematic errors are avoided or minimized (e.g., errors in the absolute T$_{\rm eff}$ scale noted above). There are several approaches to accomplishing this goal.

The California \& Carnegie group selects stars for their exoplanet survey according to magnitude ($V<8.5$), color ($B - V > 0.5$) and luminosity criteria ($M_{\rm V} > 3.0$). Thus, their survey is magnitude-limited, not volume-limited. Their control sample consists of all the stars in their survey lacking in detected planets. Complete analysis of the control sample requires considerable work with traditional analysis methods given that the survey is large (1040 stars). However, \citet{fv05} made this task manageable by analyzing the 1040 FGK stars in their survey with an automated spectroscopic method.

The Geneva group searches exoplanets in a fixed volume of space, but theirs is not a volume-limited survey. They search stars within 50 parsecs having spectral types ranging from F8 to M1 \citep{udry00}, but their survey does not include all late G, K and M dwarfs within that distance. In addition, they exclude active stars and close binaries from their survey. Their control sample is derived from those survey stars that are within 20 parsecs. Beginning with \citet{sant01} with 43 stars without known planets, their control sample has steadily grown to its present count of 93 stars \citep{gil06}.

The relative numbers of stars of different spectral types are different for the two types of surveys. A magnitude-limited survey includes relatively fewer low mass dwarfs than a volume-limited survey. \citet{fv05} provide a helpful comparison between these two types of surveys. They prepared a volume-limited test sample from their full magnitude-limited survey by selecting stars within 18 parsecs; within this distance the space density of FGK stars is constant with distance. They found that the volume-limited sample is 0.09 dex more metal-poor than their full magnitude-limited sample. They attribute the difference to the relatively larger number of metal-rich F stars in the magnitude-limited sample. Nevertheless, this difference does not account for the high mean [Fe/H] value of their SWPs subsample.

By comparing the relative incidence of SWPs in each metallicity bin, these subtle differences in the control samples are largely circumvented.

\subsection{Incidence of Exoplanets}

Using the results from their Doppler planet survey, \citet{marcy05} estimate that $>7$ per cent of nearby stars have giant planets with semimajor $<5$ AU. Extrapolating the observed distribution with semimajor axis out to 20 AU, they estimate that the incidence could be 12 per cent. 

\citet{sant05} and \citet{fv05} have calculated the relative incidence of exoplanets in each metallicity bin using their respective control sample stars. \citet{fv05} find that the incidence is 25 per cent for [Fe/H] $> +0.3$, and it is less than 3 per cent for stars with [Fe/H] $< -0.5$. They determined the following power law relation for the incidence of Doppler detected giant planets:

\begin{displaymath}
\mathcal{P} {\rm (planet)} = 0.03 \left[\frac{N_{\rm Fe}/N_{\rm H}}{(N_{\rm Fe}/N_{\rm H})_{\odot}}\right]^2
\end{displaymath}
\citet{sant04b,sant05} found a similar distribution.

\subsection{Biases}

The basic measured quantity in Doppler planet surveys is the K amplitude, which is the the radial component of the stellar orbital velocity amplitude. Anything that increases the uncertainty in the Doppler measurements will bias the observations against detection of planets producing a particular K amplitude. One such bias is S/N ratio, but planet hunters are careful to adjust the integration times to keep this quantity relatively constant. Early on, one of the planet hunting groups set their planet detection threshold at five times the S/N ratio. Now, a false alarm probability is calculated for each set of measurements of a given target star to determine a confidence level. Given the high cadence of some Doppler programs, planets are being discovered with K amplitudes approaching the S/N ratio of the data.

More importantly, the Doppler method is biased against low mass planets in long-period orbits. The stellar K amplitude assuming a circular orbit and a planet mass, $M_{p}$, much smaller than the star's mass, $M_{s}$, is

\begin{displaymath}
K = 28.4~{\rm m~s}^{-1}~P^{-1/3}~M_{p}\sin{i}~M_{s}^{-2/3}
\end{displaymath}

where $P$ is in years, $M_{p}$ is in Jupiter masses and $M_{s}$ is in solar masses.

There are also important biases dependent on the properties of the host stars. One relates to stellar age; younger stars tend to have faster rotation velocities and higher chromospheric activity. Faster rotational velocity broadens the spectral lines, yielding more uncertain Doppler shifts. Higher chromospheric activity results in greater ``velocity jitter.'' These effects have been quantified (see \citet{paul06} and references therein). F dwarfs, in particular, offer greater challenges to planet searches, since they have fewer and broader spectral lines than later spectral types. In addition, an orbiting planet of a given mass and period induces a smaller K amplitude.

In addition, bias might result from the metallicity spread among the target stars. At a given T$_{\rm eff}$, a metal-poor star will have weaker spectral lines than a metal-rich one, thus leading to more uncertain Doppler measurements. However, as argued by \citet{fv05} and as I noted in \citet{gon03}, this is not a significant source of bias among the systems studied to date. It may be a factor in a few systems with marginally detected planets.

Metal-poor stars are relatively rare in the solar neighborhood. Small number statistics for stars with [Fe/H] $\le -0.5$ for survey results published to date does limit how much we can say about planets around metal-poor stars. It also limits what we can say about the minimum metallicity required for a star to be accompanied by a giant planet. What's more, since the metal-poor tail of the distribution contains few SWPs, it is susceptible to contamination by other kinds of objects. For example, HD\,114762, with [Fe/H] $\simeq -0.6$, is sometimes included in the SWP category, but it is sometimes classified as a brown dwarf candidate, given that its minimum mass is near 12 M$_{\rm J}$. A few other brown dwarfs may have scattered into the SWP sample, but they should have a larger systematic effect on the metal-poor tail of the distribution simply because the number of metal-poor SWPs is small. \citet{sozz06} have started a Doppler survey of 200 metal-poor dwarfs ([Fe/H] $<-0.6$) in order remedy these limitations.

Finally, bias can result from the way the sample is selected. We've already noted that the present surveys are biased against very young and very metal-poor stars. More subtle biases can result from color and magnitude cutoffs. \citet{murr02} noted that a color cutoff in a survey will produce a bias against high-mass, low-metallicity stars, given the metallicity dependence of a star's color; a magnitude cutoff results in a bias against low-mass, high-metallicity stars. They concluded that neither bias can account for the observed metallicity differences between SWPs and field stars.

\subsection{Distant SWPs}

To date, six SWPs discovered with the transit method have received detailed spectroscopic analysis \citep{sant06b}. Like the nearby SWPs, these more distant SWPs are, on average, metal-rich. Most were discovered as part of the OGLE microlensing program targeting the Galactic bulge region. 

All the transit surveys directed at clusters have failed to turn up any planets. The most sensitive searches to date have targeted the globular cluster 47\,Tuc ([Fe/H] $=-0.7$). \citet{gill00} employed HST observations to search for transits in the core of 47\,Tuc; if the frequency of hot Jupiters in 47\,Tuc were the same as in the solar neighborhood, then about 17 planets should have been found in their search. \citet{wel05} used ground based observations to search the outer regions of 47\,Tuc for transits; their survey should have turned up seven planets if the incidence had been the same as in the solar neighborhood. The fact that stars in the less crowded outer regions of 47\,Tuc did not possess planets implies that metallicity, rather than crowding is the primary reason for these null results.

\section{Chemical Abundance Anomalies Among SWPs}

There have been several studies of the elemental abundance patterns among SWPs. We will review the most recent observations in this section and consider proposed explanations for the abundance anomalies in a later section.

\subsection{Light Elements}
\subsubsection{Lithium}

Stellar Li abundances are at once very informative and very difficult to interpret. This follows from the relative delicacy of Li nuclei in the shallow surface layers of stars, where they are destroyed via $(p,\alpha)$ reactions when they are mixed to regions with warm protons. Li abundances in dwarf star photospheres are observed to correlate with T$_{\rm eff}$, age, rotation, binarity and metallicity. However, even within a single open cluster (such as M67) variations in Li abundance are sometimes observed among stars that appear otherwise identical \citep{rand06}. While only one Li spectral feature is available for measurement in the spectra of sun-like stars, near 6708 \AA, it is not difficult to derive reliable Li abundances from high quality data.

Claims of Li abundance anomalies among SWPs have a complex history. \citet{gg00} suggested that SWPs, when corrected for simple linear trends with temperature, metallicity and age, display smaller Li abundances than field stars. \citet{ryan00} looked at the Li abundance trends more carefully and concluded that any possible differences are not significant. \citet{gg01}, employing a larger sample, agreed with his conclusion. 

\citet{is04} revisited this topic and reported a significant depletion in Li among SWPs relative to a comparison sample, but only in the temperature range 5600 to 5850 K. \citet{tak05} largely confirmed their findings for the temperature range 5800 to 5900 K. \citet{chen06}, restricting their attention to the temperature range 5600 to 5900 K, also confirmed \citet{is04}. However, \citet{luck06}, employing a larger sample, did not find a significant difference between SWPs and a control sample. They attribute the Li abundance difference found by \citet{is04} to a systematic difference in the temperature scales in their study and the study of \citet{chen01}, the results of which they had used to supplement their comparison sample. Clearly, this topic should be revisited as the number of SWPs in the 5600 to 5850 K temperature range continues to increase.

\citet{israel01} reported the presence of $^{6}$Li in the atmosphere of HD\,82943, a metal-rich SWP; They measured a $^{6}$Li/$^{7}$Li ratio of $0.126\pm0.014$. \citet{redd02} searched for $^{6}$Li in 6 field stars and 8 SWPs, including HD\,82943, without success. \citet{is03} redetermined the $^{6}$Li/$^{7}$Li ratio of HD\,82943 to be $0.05\pm 0.02$, using a higher quality spectrum and an updated linelist for the Li region. Employing yet another modified linelist, \citet{man04} determined the $^{6}$Li/$^{7}$Li ratios for three SWPs and two comparison stars, finding only upper limits. While they did not include HD\,82943 in their study, their results are consistent with \cite{redd02} for the stars in common.

\subsubsection{Beryllium}

Studies of nearby sun-like stars have shown that Li is depleted more rapidly than Be in their atmospheres, largely confirming theoretical expectations that higher temperatures are needed to destroy Be \citep{boes04}. \citet{sant04} determined Be abundances in 41 SWPs and in a comparison sample of 29 stars without known planets. They did not find a significant difference between the two groups.

\subsection{Other Elements}

About 20 other elements have been included in the larger abundance studies of SWPs. In this section I will summarize results of such studies published since 2001 that include abundances of multiple elements (for a review of older studies, see \citet{gon03}). Note that comparisons of the abundances of SWPs and stars without known planets are usually made relative to Fe, as [el/Fe]; doing so removes the already known difference in [Fe/H] between the two groups.

1 - C, N, O: The abundances of these elements are somewhat challenging to determine with precision. These elements are represented by relatively few weak lines in the spectra of FGK stars. The exception is the O I triplet at 7770 , but O abundances determined from them require non-LTE corrections. \citet{gg01} compared the [C/Fe] and [O/Fe] values of 38 SWPs to those of several field stars samples from the literature and did not find any significant differences. \citet{takhon05} did not find any significant differences in [C,N,O/Fe] values between their samples of 27 SWPs and 133 control stars. Neither did \citet{luck06} find any differences in [C, O/Fe] between their samples of 55 SWPs and 161 control stars.

2 - Na, Mg, Al: These elements are also represented by relatively few absorption lines. \citet{gg01} found slightly lower values of [el/Fe] for these three elements among SWPs. \citet{bei05} measured these elements in 98 SWPs and 41 comparison stars and failed to find significant differences. However, the same group, expanding the samples to 101 SWPs and and 93 comparison stars, found lower values for [Al/Fe] and higher values of [Mg/Fe] among the SWPs relative to the comparison stars \citep{gil06}. \citet{luck06} did not find differences between their 55 SWPs and 161 comparison stars.

3 - Si, S, Ca, Sc, Ti: Of the elements in this group, S is the most difficult to measure in spectra; only recent determinations using high dispersion, high S/N spectra should be trusted. [Si/Fe] is usually determined with higher precision and displays less scatter than other element abundance ratios. Employing sophisticated bootstrapping statistical analysis, \citet{rob06} found that the 99 SWPs in their Spectroscopic Properties of Cool Stars (SPOCS) database have significantly higher [Si/Fe] values than the 941 stars in their control sample.

4 - V, Cr, Fe, Ni, Co, Zn, Cu: \citet{rob06} also found higher [Ni/Fe] values for SWPs relative to comparison stars in their SPOCS database. Determinations of V abundances often display large scatter, given that the transitions producing the measured lines are low-excitation and hence sensitive to errors in temperature. \citet{gil06} found differences between SWPs and their control sample for V and Co. Accurate Cu abundances requires spectrum synthesis and inclusion of hyperfine components; this has been done in the recent studies.

5 - heavy elements: Some elements beyond the iron-peak, such as Ba and Eu, have been measured in SWPs. \citet{luck06} have measured the greatest number of heavy elements in SWPs, but they did not find any significant anomalies.

Summarizing published research, there is some evidence that SWPs differ from other nearby FGK stars without planets in their abundances of Mg, Al, Si, V, Co and Ni. 

\subsection{Common Proper Motion Pairs}

Common proper motion (CPM) pairs are binary stars with large separation, usually discovered in astrometric proper motion surveys. Stars in CPM pairs presumably formed together out of the same birth cloud and should have the same initial composition (except for Be and Li). Those selected for study are sufficiently separated on the sky so that a spectrum of each component can be obtained without contamination from its companion. 

The 16 Cyg system was the first CPM in which a planet was found (around the B component). It had been known for several years that the two stars have significantly different lithium abundances, yet similar effective temperatures. \citet{lawgg01}, applying a high-precision differential abundance analysis spectroscopic technique, reported a significant difference in [Fe/H] between two stars. \citet{tak16}, employing a refined version of their method, found, instead, that [Fe/H] is the same in the two stars to within 0.01 dex.

Employing a similar differential analysis method, \citet{grat01} reported a significant metallicity difference between the pair of stars in the HD\,219542 binary. At the time, they had also reported a planet in orbit about the B component. The group later retracted both claims in follow up studies. To date, they have analyzed the spectra of 50 CPMs \citep{desi04,desi06}. One pair in their sample, HD\,113984, displays a difference in [Fe/H] of 0.27 dex, but the primary is a blue straggler. This allows for the possibility that the blue straggler's anomalous [Fe/H] value is a result of the formation process of the blue straggler. A few other pairs display [Fe/H] differences between 0.05 and 0.09 dex, but the authors are tentative in their conclusions. They suggest that diffusion is the best explanation for them.

\citet{luck06} included nine CPM pairs in their spectroscopic survey and found two with significantly discrepant compositions. One CPM pair, HR\,7947/HR\,7948, differ in [Fe/H] by 0.25 dex, though neither is known to host a planet. The other pair is HR\,7272A/B. The B component has a planet, and its metallicity is 0.2 dex larger than that of the A component. The A and B components also have very different Li abundances.

In summary, there is intriguing evidence that some CPM pairs have different compositions, but these need to be confirmed.

\subsection{Transiting Planets}

Spectroscopic analyses of stars with transiting planets show that they are metal-rich (\citet{sant06c,guil06}). Transiting planets offer the opportunity to determine the radii and densities of short period planets, something not possible with other methods. From a sample of 9 stars with transiting planets, \citet{guil06} found that the heavy element content of the planets correlates with the metallicity of the host stars. \citet{sant06c} found that stars with the shortest period planets are the most metal-rich. In both cases the authors caution that their results are preliminary.

\section{Possible Causes of Compositional Anomalies}

\citet{gon03} summarized three classes of causes/explanations for the observed compositional anomalies among SWPs: self-enrichment; migration; primordial. Briefly, self-enrichment refers to significant alteration of a star's surface composition by the accretion of metal-rich material. For the migration explanation, I am substituting in the present review a broader idea, ``orbital period bias''. The orbital period bias explanation posits that the orbital radius (and hence period) of a giant planet depends on metallicity. The primordial explanation posits that the probability of forming giant planets depends on the initial metallicity of a system's birth cloud. I present descriptions of these three explanations and ways to test for them below, followed by evaluations of recent studies.

\subsection{Self-enrichment}

By the time the Sun was a few hundred million years old, its outer convection zone had shrunk to about 0.03 M$_{\odot}$. For a 1.2 M$_{\odot}$ star, the outer convection zone is only 0.006 M$_{\odot}$ at the same age. Thus, the sensitivity of the surface metallicity to accreted matter rises steeply for stars only slightly more massive than the Sun. The addition of half an Earth mass of iron to the Sun would have increased its [Fe/H] by about 0.017 dex (\citet{murr01}; see also figure 5 of \citet{ford99} for estimates of the increase of [Fe/H] in the solar atmosphere for various amounts of added rocky material).

There is no doubt that the Sun has accreted metal-rich material; even today comets are observed colliding with it. The processes involved in planet formation virtually guarantee that stars accompanied by planets will experience such accretion. \citet{jeff97} estimated the early Sun could have accreted as much as $100_{\oplus}$ of metal-rich material. This much accretion by the early Sun is ruled out by solar models \citep{gon06}, assuming it occurred  late enough that its convection zone was already shallow. A more modest $25_{\oplus}$ of accreted material would result in metallicity increase of 0.14 dex, an amount comparable to the average metallicity difference between SWPs and control samples. 

Another possible complication is penetration of the core of a giant planet through a star's convection zone. \citet{sand02} argued that this could happen if a ``cold'' giant planet plunges into a star slightly more massive than the Sun. Thus, the quantity of material accreted in a stellar convective envelope depends on its nature (rocky versus giant planet) and its orbital dynamics.

There are several possible observational tests of the self-enrichment hypothesis. One approach is to search for a trend between [Fe/H] and stellar convection zone mass. F dwarfs have less massive convection zones than G and K dwarfs and, thus, will exhibit greater metallicity enhancement for a given accreted mass. Care needs to be taken that the generally younger ages of F dwarfs it taken into account. Old metal-poor F dwarfs no longer exist; they evolved off the main sequence long ago.

In addition, once a star leaves the main sequence, its convection zone deepens dramatically. If a star's envelope was enriched on the main sequence, then the metals in its envelope will be diluted once it ascends the subgiant brach. 

A trend of element abundance with condensation temperature, T$_c$, is another possible test of self-enrichment. Material that condenses closest to the host star will be most depleted in volatile elements. Given this, accretion of condensed material onto the star will preferentially enrich it in elements with high T$_c$. A star that has experienced accretion of high-T$_c$ material will display an excess positive correlation between element abundances and T$_c$ compared to a control sample. As noted by \citet{gon03} these two observational tests can be combined; F dwarfs should display steeper trends with T$_c$ than G and K dwarfs if accretion has occurred. 

\citet{alex67} was the first to suggest that accretion of a planet onto a star could significantly increase its surface Li abundance. This follows because a star like the Sun gradually depletes the Li in its convective envelope, while planets preserve their Li. However, attributing the measured Li abundance in a star to accretion is not straight-forward. The physical properties of the star, such as its age and metallicity, need to be well-determined in order to correct for the processes that alter Li in its envelope.

Less ambiguous evidence for self-enrichment in a star would be the detection of $^{6}$Li in its atmosphere. A metal-rich dwarf later than mid-F destroys all its primordial $^{6}$Li by the time it reaches the main sequence while it retains much of its $^{7}$Li \citep{mont02}. 

Finally, asteroseismological observations can be used to test enriched stellar models against homogenous  composition models \citep{baz05}. This is the only direct way to constrain observationally the internal composition of a star other than the Sun. The best targets for this type of test are nearby old F and G dwarfs; in addition to the measurements of the acoustic p-modes, it is important to have accurate measurements of luminosity, T$_{\rm eff}$ and surface [Fe/H].

\subsection{Orbital Period Bias}

The Doppler and transit planet search methods are strongly biased in favor of planets with small orbits. Any physical processes that result in planets with small orbits will enhance their detectability with these methods. One such process is planet migration. If migration is sensitive to the initial metallicity, such that more metal-rich systems are more likely to experience inward migration of the giant planets, then a survey will show a correlation between the presence of giant planets and the metallicity of the host star. One way to test for this bias is to search for a correlation between planetary semimajor axis and host star metallicity. 

Since planet migration might be linked with self-enrichment \citep{mural02}, before the signature of migration can be detected, it will be necessary to identify systems that have been enriched. It is not clear, however, how to treat multiple-planet systems, and some stars presently known to have only one giant planet may later be found to have additional giant planets in larger orbits.

It could be the case that all the extrasolar giant planets known to date are ``short-period'' systems that have experienced substantial migration. Perhaps there is a group of systems with giant planets like Jupiter that have not experienced substantial migration. \citet{rice05} suggest that systems that have not undergone migration could be identified with the lower envelope of a plot of metallicity versus semi-major axis. They argue that at any radius, giant planets that barely manage to form around the lowest metallicity stars on the plot accrete their gas envelopes just as the disk gas is dissipating. Such stars should experience little or no migration.

\subsection{Primordial}

Support for the primordial explanation is motivated by the CIA model of giant planet formation. In brief, a higher initial metallicity results in a higher density of solids in a protoplanetary disk, which, in turn, increases the probability that giant planet cores will form before the disk gas is lost. This model predicts that the incidence of giant planets around sun-like stars should increase with increasing metallicity.

Confirmation of the primordial explanation is primarily a process of elimination of the other two classes of explanations. First, the importance of self-enrichment must be determined and the sample of SWPs corrected for its effects. Second, any biases related to metallicity must be characterized and corrected for. Third, any remaining trends with metallicity will be attributable to the primordial explanation.

\subsection{Evaluation}

Most of the tests described above are not new; some were proposed as early as 1997. Several have been applied to SWPs multiple times. However, a failed test for a given sample does not mean that the test should be abandoned. If the phenomenon being tested, such as self-enrichment, is stochastic and rare, then a large sample size is necessary before a convincing instance of it is found. Even a null result can be useful in constraining planet formation models.

Much evidence has been cited in favor of the self-enrichment explanation. For a time, measurement of an anomalously high $^{6}$Li abundance in the SWP HD\,82943 seemed convincing. However, as noted above, this result is now in dispute. While observational evidence for $^{6}$Li in some metal-poor stars has been reported in the literature \citep{asp06}, it has been attributed to a pre-Galactic origin. 

Evidence for compositional differences between CPM pairs is also weaker than what was originally reported. The case for HD\,219542 has fallen apart, and the case for 16\,Cyg has been challenged. The two new cases discussed by \citet{luck06}, however, deserve further attention. 

\citet{paul03} failed to find small metallicity variations among secure Hyades cluster members. This null result is important, given that the Hyades is a metal-rich cluster, and its member stars are expected to have a high incidence of giant planets. \citet{shen05} also searched for evidence of enrichment in IC 4665, a young, solar-metallicity open cluster. They failed to detect significant correlations between [Fe/H] and convection zone mass or between [Fe/H] and Li abundance.

\citet{takbig06} searched for trends with convection zone mass in the large SPOCS database. They noted that the most metal-rich SWPs are G and K dwarfs, not F dwarfs as would be expected from self-enrichment. They also note that of the 10 subgiants with planets in SPOCS, three (HD\,27442, HD\,38529, HD\,177830) are members of the rare super metal-rich class. Both these findings are inconsistent with self-enrichment being the source of the high metallicities of SWPs.

\citet{smith01} was the first study to search for anomalous trends between chemical abundances and T$_{rm c}$ among SWPs. They calculated the abundance-T$_{\rm c}$ slope for a sample of SWPs and compared them to a control sample. After correcting for a trend due to Galactic chemical evolution, they noted that six SWPs have anomalously high T$_{\rm c}$ slope values. Several recent studies have revisited this test \citep{ecu06, gon06, huang05} and concluded that there are no significant differences between SWP and control star samples.

\citet{murr02} compared the [Fe/H]-stellar mass trends in a sample of 50 SWPs and a control sample of 466 stars. They concluded that the average SWP has accreted 6.5 M$_{\oplus}$ of Fe. One of the evidences they cited in support of this conclusion is the steeper slope between mean [Fe/H] and mass for the SWPs compared to the control stars. 

It is worthwhile to revisit \citet{murr02}'s test for self-enrichment with the larger SWP and control star samples of \citet{fv05}. We show in Figure 1 binned data for the SWPs and the control stars from their work (highly evolved subgiants were removed from the samples by requiring that $\log g \gtrsim 3.8$). The increase of [Fe/H] with mass in the two samples is due to the stellar age-metallicity relation (older stars are more metal-poor) combined with the fact that more massive stars have shorter lifetimes. It is clear that the two groups of stars have indistinguishable slopes. Unfortunately, the range of main sequence masses for which \citet{murr02} predict the most dramatic rise of [Fe/H], 1.4 - 1.8 M$_{\odot}$, is not yet in the range of Doppler planet surveys.

\citet{murr02} cited the smaller mean [Fe/H] values of ``Hertzsprung gap'' stars relative to less evolved stars as further evidence for enrichment (in this case, the purported enrichment is not specifically connected with the presence of planets). They define a Hertzsprung gap star as having at least 10 times the mass of the convection zone it had on the main sequence; Hertzsprung gap stars are a subset of  subgiants. A Hertzsprung gap star should have thoroughly diluted any accreted metals in its envelope from mixing with its deeper layers. \citet{murr02} also found that the [Fe/H] values of their 19 Hertzsprung gap stars have a weaker dependence on mass than their sample of dwarf stars. According to their best-fit enrichment model the average dwarf star has accreted 0.4 M$_{\oplus}$ of Fe in its envelope.

Employing the larger samples from \citet{fv05}, we do confirm that Hertzsprung gap stars are more metal poor than less evolved stars. In contrast to \citet{murr02}, our sample of 34 Hertzsprung gap stars displays a steep correlation between [Fe/H] and mass (Figure 1; in selecting the Hertzsprung gap stars from the database of \citet{fv05}, we made use of the convection zone mass estimates of \citet{takbig06}). Care should be exercised in interpreting these trends, however. Subgiants are more evolved and, therefore, older than dwarfs of the same mass. The progenitors of subgiants with masses just below the Sun's, in particular, have lifetimes comparable to the age of the Galaxy. It will be necessary to run simulations that take into account Galactic chemical evolution before we can conclude that Hertzsprung gap stars have anomalously small metallicities compared to dwarf stars. Another approach is to remove age as a variable by comparing evolved and unevolved stars in a cluster. \citet{rand06} is the first such study; they did not find any significant differences between dwarfs and subgiants in M67.

Following up on earlier reports, \citet{sozz04} confirmed that SWPs with hot Jupiters (specifically, orbital period, $\it P$, less than about 5 days) are more metal-rich than the mean of other SWPs. The mean [Fe/H] value of the 14 SWPs with short-period planets listed in his study is 0.22. \citet{but06} list an additional 14 SWPs with short-period planets; of these, 11 have known [Fe/H] values. The mean [Fe/H] value of the 25 SWPs with short-period planets is 0.18. Three of them have [Fe/H] values less than 0; the lowest value of [Fe/H] is -0.18, sill about 0.45 dex larger than the lowest known value of [Fe/H] in the full sample of SWPs.

The more homogeneous database of \citet{fv05} also lends support to the conclusions of \citet{sozz04}. The 15 SWPs with planets in orbits with $\it P$ $< 5$ days in \citet{fv05} have a mean [Fe/H] of 0.23, while the remaining 89 SWPs have a mean [Fe/H] of 0.12. None of the SWPs with short period planets have [Fe/H] values below solar. It is notable that the mean value of [Fe/H] for the SWPs with planets having $\it P$ $> 5$ days is still significantly larger than that of the control sample.

While \citet{sozz04} attributes the high [Fe/H] values among SWPs with short period planets to metallicity dependent planet migration, \citet{pin05} propose a model wherein a giant planet's place of formation depends on metallicity. Their model predicts that the most probable place of formation of a system's most massive planet shifts closer to the host star for lower metallicity. Migration is then more likely to result in planet engulfment and, hence, leaving the system without a planet to be detected. This model could be tested by searching for evidence of self-enrichment among metal-poor stars.

The strongest evidence for self-enrichment comes from analysis of a star without known planets. \citet{lawgg03} and \citet{ash05} argue that the chemical abundance pattern of the early F dwarf J37 in the open cluster NGC\,6633 is best explained by accretion of rocky material. This star was first noticed on account of its exceptionally high Li abundance (an order of magnitude above the meteoritic value). Its chemical abundances display a positive correlation with T$_{\rm c}$. It is not known if J37 is accompanied by planets. 

A more exotic example of self-enrichment is described by \citet{jura06}. Citing the case of J37, he argues that the low measured C/Fe abundance ratio in the atmospheres of three white dwarfs is best explained by accretion of circumstellar asteroidal material, rather than interstellar material.

\citet{baz04} and \citet{baz05} compared normal and self-enriched stellar model p-mode predictions to asteroseismological observations of the metal-rich SWP, $\mu$ Arae. Their metal-enriched stellar model gives a modestly better fit to the observations than their homogeneous model.

Galactic kinematics can help discriminate among the three classes of explanations for the correlation between metallicity and the presence of planets. If only the primordial explanation is important, then giant planets will form once some critical minimum metallicity is exceeded, regardless of the Galactic kinematics. \citet{bar02} and \citet{sant03} did not find any differences between the kinematics of SWPs and a control sample. Employing a larger sample of SWPs, \citet{ecu06c} found that the SWPs sample is more similar to their metal-rich comparison subsample than it is to the full comparison sample. They interpret this as supporting the primordial explanation. They speculate that most SWPs were formed interior to the solar circle (where more metal-rich stars form) and migrated outward to their present Galactic positions. However, \citet{bar02} noted a trend in the lower envelope of the [Fe/H] versus perigalactic distance plot for SWPs. They interpreted this as favoring the self-enrichment hypothesis. Overall, the findings from Galactic kinematics studies to date appear to be ambiguous.

In summary, only weak evidence exists for self-enrichment among the known SWPs. There is some evidence that SWPs with short-period planets are more metal-rich than SWPs with long-period planets, but there are several possible explanations for this. The bulk of the metallicity dependence on giant planet incidence is best explained by the primordial hypothesis.

\subsection{Implications for Models of Planet Formation}

Characterization of the correlation between the composition of the atmosphere of a star and the presence of giant planets gives us helpful constraints on planet formation and evolution processes. In particular, the fact that the evidence reviewed above favors the primordial explanation lends support to the CIA model of planet formation.

Assuming the CIA giant planet formation model and the primordial explanation, \citet{idalin05} were able to reproduce the observed distribution of giant planets with host star metallicity. They also predict that close-in giant planets should be rare around M dwarfs, but neptune mass planets should be common around them. The former prediction already has some observational support \citep{laws03,endl06}.

\citet{rice05} argue that if metallicity is the dominant factor controlling the timescale for giant planet formation, then there should be a correlation between planet mass and metallicity. They show that the data confirm this prediction, but the statistical significance is only modest. Certainly, this prediction deserves revisiting as the SWP sample size continues to grow.

If self-enrichment is implicated in some cases, then we may be able to determine the composition of the accreted material. It is almost axiomatic that $\it{some}$ accretion of rocky material onto a star will occur if it is accompanied by a disk containing condensed particles. It now appears from the latest data that the chemical anomalies resulting from self-enrichment are only detectable among main sequence dwarfs earlier than about mid-F spectral type and cool white dwarfs.

\section{Recommendations for Research and Concluding Comments}

The following list of suggested research topics may help advance our understanding of planet formation and evolution by studying compositional anomalies among SWPs. While most are not new suggestions, they need to be pursued more fully than they have been. They are arranged roughly in order of decreasing priority/increasing difficulty.

The most firmly established compositional anomaly among SWPs is their high mean metallicity relative to control samples. The search is now on for more subtle anomalies. As shown by several recent studies (e.g., \citet{luck06,bond06,gil06}), mean compositional differences between SWPs and control samples, expressed as [X/Fe], must be smaller than about 0.15 dex. It is possible to detect compositional anomalies using current techniques even at this level, but it will be necessary to control carefully for systematic errors.

Apart from the controversies surrounding Li abundances, only \citet{rob06} have claimed to discover statistically significant compositional anomalies among SWPs. Their findings need to be confirmed by independent analyses. Continuing spectroscopic study of nearby SWPs stars may confirm the other compositional differences suggested by some studies (e.g., Al/Fe). Addition of more metal-rich ([Fe/H] $> 0.3$) comparison stars will also be helpful in this regard. It is especially valuable to have several independent research groups testing each other's findings, given the subtleties of the claimed anomalies and the possibility of unaccounted for systematic errors (e.g., \citet{schul06}). The recent history of research on this topic has shown on multiple occasions that findings that originally looked convincing were not confirmed following additional research.

On the theoretical front, published abundances of SWPs and control samples should be corrected for departures from LTE. Presently, only SWP O abundances derived from the O I 777 nm triplet lines include such corrections. The differential (relative to the Sun) corrections for most published abundances of these stars are expected to be considerably less than 0.1 dex. Studies could be modeled after \citet{idiart00}, for example, which calculated non-LTE corrections to Ca and Mg abundances using published EW values. Another way to improve the abundances determinations is to replace the 1D stellar model atmospheres with 3D model atmospheres \citep{asplund05}. These corrections are especially needed now that planet hunters are expanding their searches to stars outside the original dwarf FGK range.

Application of these corrections to spectroscopic analyses of FGKM dwarfs can be facilitated by setting up a grid of standard stars spanning the same range of spectral types. A small number of standard stars could be analyzed with the most recent 3D atmospheres and non-LTE calculations using time-consuming methods. The new standards could then be employed to correct chemical abundances of stars analyzed using the simpler traditional methods. For best results, the analyzed stars would be compared to standards closest in spectral type and metallicity. The standard stars could be analyzed again with every new generation of model atmospheres and analysis methods.

Several CPM pairs have been reported to display significant differences in [Fe/H]. These need to be confirmed with independent analyses. Proper interpretation of the results will require consideration not only of self-enrichment but also of diffusion. Both diffusion and self-enrichment are expected to become more important for hotter dwarfs. The two processes can be separated if a sufficient number of elements is included in the analyses \citep{ash05}.

It would also be worthwhile to continue the search for signatures of self-enrichment among early type stars. In particular, the strong evidence for self-enrichment in J37 in NGC\,6633 serves as motivation to conduct a general spectroscopic survey of similar stars in clusters of comparable age and metallicity. More generally, studies similar to \citet{dot03}, which searched for trends between [Fe/H] and spectral type and between the scatter in [Fe/H] and spectral type in Hyades A and F stars, should continue to be refined and applied to additional clusters.

As noted recently by \citet{jura06}, the atmospheres of cool helium-rich white dwarfs are sensitive to small amounts of accretion. Determination of the abundances of additional elements in their atmospheres should permit more stringent tests of the two competing explanations for the anomalous cases: interstellar and circumstellar accretion. 

While attempts to detect the signature of self-enrichment via abundance-T$_{\rm c}$ trends have not proven successful, this test should continue to be applied periodically as the database of SWP and control sample abundance determinations continues to grow. Although the test can be refined by any improvements to abundance determinations, the greatest benefits will come from improvements for elements with low values of T$_{\rm c}$, such as C and O. O abundances, in particular, require careful attention, given recent controversies \citep{ecu06b}. If self-enrichment is a rare process, then evidence for it will be seen in a small number of extreme stars, like J37. Otherwise, it could be detected as a statistically significant trend in a large sample. Regardless of which is the case, any analysis of this type should include as many F dwarf stars as possible.

Comparison of the metallicities of dwarfs and subgiants in a cluster will allow us to test self-enrichment much more effectively than similar studies of field stars. An analysis similar to that of \citep{murr01} could be conducted on an old, solar metallicity open cluster with a well-populated turn-off and subgiant branch (e.g., M67). Stars that had been enriched on the main-sequence should display reduced surface metallicity on the subgiant branch. Such a study would avoid the weakness of \citet{murr01}'s less homogeneous sample. An even better approach would be to compare enriched stellar evolutionary tracks \citep{cody05} to star cluster color-magnitude diagrams.

As noted above, some planet formation models proposed to explain the metallicity-planet link also predict certain trends with stellar mass. Together, these two stellar parameters more tightly constrain such models than either of them alone. Planet surveys that target stars outside the main sequence FGK spectral type range, such as \citet{gal06}, \citet{endl06} and \citet{john06}, will be especially helpful in testing the possible stellar mass-planet link.
 
To date, only one SWP has been targeted for asteroseismological analysis, $\mu$ Arae \citep{baz04,baz05}. The analysis results tend to support the self-enrichment model for this star, but additional data are needed to reach a firm conclusion. \citet{baz05} suggest that interferometric radius measurements of $\mu$ Arae accurate to better than 1 per cent would provide a significant new constraint on the models. Significant improvements in p-mode identifications would require a substantial new Doppler observing campaign.

Doppler planet searches in open clusters have not yet proven successful, but only one cluster, the Hyades, has been targeted. Focusing on an open cluster removes age and initial metallicity as variables. The Hyades is the closest cluster, and it is metal-rich, but its stars are apparently too young to allow detection of planets with confidence \citep{paul04}. It will be necessary to target clusters comparable in age and metallicity to the Sun, such as M67. Doppler searches for planets in clusters beyond the Hyades will require at least 8-m class telescopes. Allocation of the needed large telescope time for these searches may have to await better estimates of the expected detection efficiencies from open cluster transit surveys.

\subsection{Concluding Comments}

The composition of the typical nearby SWP differs from that of the typical nearby star. Several recent studies have established with a high degree of statistical confidence that, as a group, SWPs are more metal-rich than stars without known planets. The data are also suggestive of differences in the Al/Fe, Si/Fe and Ni/Fe ratios. Progress in testing the reality of these differences can be made by including more nearby stars without planets having [Fe/H] $> 0.3$ in the control samples.

The primordial hypothesis is the best single explanation for the anomalously high metallicities of nearby SWPs. The data are also suggestive of an orbital period bias, implying a metallicity-dependent mechanism that affects the final orbital position of a planet. However, an updated analysis of its statistical significance is required. Evidence for self-enrichment among SWPs is still weak, but there is the expectation that convincing evidence for it will be found among early F dwarf SWPs discovered in ongoing Doppler planet surveys.

Although the number of planets detected with the transit method is still small, it is already evident that their host stars are metal-rich. There is also evidence for a correlation between the metallicities of the host stars and the planets of transiting systems.

Taken together, these results favor the CIA model of giant planet formation. While some progress has been made in explaining the composition of SWPs within the framework of the CIA model, much remains to be done. The competing DGI model may still account for a small fraction of the known planets. Its contribution could be quantified once the CIA model is better constrained and the number statistics for metal-poor SWPs is improved.

\acknowledgments

I thank the anonymous reviewer for helpful comments and suggestions that resulted in a much improved revision of the origin version of this review.

\clearpage

\begin{figure}
%\plotone{f1.eps}
\caption{[Fe/H] versus stellar mass for 104 SWPs (filed circles), 788 control stars (open circles) and 34 Hertzsprung gap stars (triangles), binned in intervals of $\sim 0.1$ M$_{\odot}$. The data are from \citet{fv05}. See text for additional details.}
\end{figure}

\end{document}